\documentclass[
showpacs,%floatfix,
aps,prl,amsmath,amssymb,
twocolumn,floatfix
]{revtex4}

\usepackage{psfrag}

\usepackage{graphicx}

\begin{document}

\title{Generic features of the spectrum of trapped polarized fermions 
}

\author{Erich J. Mueller}
\affiliation{Laboratory of Atomic and Solid State Physics, Cornell University, Ithaca NY 14853}

\pacs{03.75.Ss%Degenerate Fermi gases(LB, GR+andreev,
,05.30.Fk%Fermion systems and electron gas(Levitov-barankov(LB),
%,03.75.Kk%Dynamic properties of condensates; collective and hydrodynamic excitations, superfluid flow (LB,
%,03.75.Hh%Static properties of condensates; thermodynamical, statistical, and structural properties (we probably dont want this)
%,71.10.-w %Theories and models of many-electron systems
%,05.30.Jp %Boson Systems (one of Gurarie-Radzihovsky(GR) papers had it.)
}

\begin{abstract}
We show that bimodal radio frequency spectra universally arise at intermediate temperatures in models of strongly interacting trapped Fermi gases.  The bimodality is independent of superfluidity or pseudogap physics, depending only on the functional form of the equation of state -- which is constrained by dimensional analysis at low temperatures and the  virial expansion at high temperatures.  In addition to these model independent results, we present a simple calculation of the radio frequency line-shape of a highly polarized Fermi gas which uses energetic considerations to include final state interactions.  While this model only qualitatively captures the line-shapes observed in the experiments, it provides a conceptually clean and powerful technique for estimating the energy scales and how they vary with experimental parameters.
\end{abstract}
\date{\today}
\maketitle

Can spectroscopy be used to detect pairing in a gas of fermionic atoms?  The paradigm for thinking about this question was set in 2003, when Greiner, Regal and Jin \cite{Jin-2005} presented an experimental study of the microwave spectrum of a two component gas of $^{40}$K on the BEC side of resonance (where two potassium atoms are capable of forming weakly bound molecules).  Their sample consisted of a noncondensed gas of diatomic molecules in chemical  equilibrium with a gas of atoms.  They found an easily interpreted bimodal spectrum: a sharp line came from the excitations of free atoms, and a broad peak from the dissociation of molecules.  Later experiments on colder samples in different parameter ranges showed similar bimodality and were interpreted in similar ways \cite{chingap,ketterlepairs}.  In particular, when such a bimodal peak was seen by Chin, Bartensein, Altmeyer, Riedl, Jochim, Denschlag, and Grimm \cite{chingap} on the BCS side of resonance, it was taken as a sign of Cooper pairing.  This interpretation was further reinforced by mean field calculations which showed that the finite temperature trapped paired gas does indeed produce bimodal spectra \cite{Torma-2004,Ohashi-Griffin-2005,Levin-2005,Levin-2007}.  This paradigm has been shattered by the realization that there exist models which produce bimodal spectra in the absence of pairing \cite{massignan}.  Here we give a simple argument for why trapped strongly interacting Fermi gases {\em generically} show a bimodal spectrum, irrespective of the presence of pairing.

Our argument relies on two ingredients: a harmonic trap plus qualitative features of the equation of state.  
We establish that the equation of state of the unitary Fermi gas has these features through a dimensional analysis argument supplemented by the first terms of the virial expansion.   Consistent with previous results \cite{massignan,Torma-2004,Ohashi-Griffin-2005,Levin-2005,Levin-2007}, we conclude that one of the spectral peaks comes from atoms at the edge of the cloud, and the other from atoms at the center.  This is similar to the mechanism by which multiple spectral peaks appear in the spectra of  clouds of Bosons in an optical lattice \cite{ketterlespec,hazzardmueller}.

We quantify our arguments
by presenting a simplified calculation of the spectrum of a 
%harmonically 
trapped two-component Fermi gas in the limit  $n_\downarrow/n_\uparrow\to0$.
Although we include final-state interactions \cite{Baym-2006,Zwerger-2007,Baym-2007,perali-RF,Levin-2008,newstoof,Chin-Julienne-2005,sheehy,basumueller} in our quantitative calculations, we emphasize that the bimodality is {\em not} a final-state effect.  In the  highly polarized limit, final state effects set the frequency scale for the spectral line but do not significantly change its shape or temperature dependence.  Moving away from this limit of large polarizations, final state interactions can have more profound effects \cite{Baym-2006,Zwerger-2007,Baym-2007,perali-RF,Levin-2008,newstoof,Chin-Julienne-2005,sheehy,basumueller,insensitive,Ketterle-II-2007,Ketterle-2008}.

We consider an experiment on a gas of atoms with two spin state: $\uparrow$ and $\downarrow$.  We imagine that a probe (for example radio waves) excites atoms from the $\downarrow$ state to a third state $x$.  In most recent experiments on $^6$Li the states involved are the three lowest energy hyperfine states $|\!\!\uparrow\rangle=|1\rangle$, $|\!\!\downarrow\rangle=|2\rangle$ and $|x\rangle=|3\rangle$, ($E_1<E_2<E_3$) though other combinations are possible \cite{Ketterle-2008}.  The spectrum $I(\nu)$ measures, for a fixed probe intensity, the rate of population transfer from $\downarrow$ to $x$ as a function of the detuning $\nu$ from the free-space resonance.  

In principle, moments of this spectrum can be calculated from the sum rule \cite{Baym-2006,Baym-2007,Zwerger-2007},
$\bar\nu$$=$$\int d\nu \nu I(\nu)/\int d\nu I(\nu)$ $=$ $\frac{1}{n_\downarrow}\int d^3r$ $\left[v_{\uparrow x}(r)-v_{\uparrow \downarrow}(r)\right]$ $\langle \psi_\uparrow^\dagger(r)\psi_\downarrow^\dagger(0)\psi_\downarrow(0)\psi_\uparrow(r)\rangle$, where $v_{ij}(r)$ is the interaction potential between atoms in states $i$ and $j$, and $\psi_\sigma$ is an annihilation operator.
As pointed out by Pethick and Stoof \cite{pethstoof}, extracting useful information from this result is problematic because this expectation value (which is closely related to the expectation value of the interaction energy) is not a low energy observable: different potentials which give rise to the same low energy scattering properties will give completely different values for $\bar \nu$; hard spheres  have $\bar \nu=0$, while point interaction yield $\bar\nu=\infty$.  These potential dependent features come from the largely unobservable ultraviolet tail of the spectrum.  The peak of the spectral line occurs at relatively small $\nu$ and only depends on low energy properties such as the scattering lengths $a_{ij}$.

  Here, we explicitly restrict our discussion to the peak of this extended spectrum, beginning with the conceptually
 simple limit where there is a single down-spin particle in a gas of up-spin particles with {\em uniform} density $n_\uparrow$.  Barring pathologies in the matrix elements of the probe, and subtleties associated with vertex corrections \cite{pethstoof}, the peak in $I(\nu)$ should be located at detuning 
 \begin{equation}
 \nu_p(n_\uparrow)=\epsilon_x(n_\uparrow)-\epsilon_\downarrow(n_\uparrow),
 \end{equation}
  where $\epsilon_\sigma(n_\uparrow)$ is the energy of a single particle of spin $\sigma$ in the sea of $\uparrow$ atoms.  Note that $\epsilon_\sigma$ is neither the self-energy, nor the chemical potential, though it can %in principle
   be derived from the self-energy by integrating with respect to the coupling constant \cite{mahan}.  The fact that $\epsilon_\sigma\neq \mu_\sigma$ is clear from considering the dilute high temperature limit, where $\mu_\sigma$ is very negative, but $\epsilon_\sigma$ is very small.  Assuming that the inhomogeneous broadening dominates the lineshape, one can approximate the homogeneous lineshape as a delta function.  In a trapped gas, applying this approximation locally will yield an inhomogeneous spectrum
\begin{equation}
I(\nu)\propto \int d^3 r\,n_\downarrow(r) \delta\left(\nu-\nu_p(n_\uparrow(r))\right).
\end{equation}   
   Note that we are explicitly neglecting the role of ``Frank-Condon" factors, related to overlaps between the states in the two channels, and do not include ``excitonic" effects related to pairing in the final  state   \cite{perali-RF, basumueller,sheehy,newstoof,Levin-2008}.  To correctly model the uniform spectrum, as experimentally studied in \cite{Ketterle-II-2007,Ketterle-2008}, it would be essential to include such factors.
% In particular, the considerations of Fumarola and Mueller \cite{fumarolamueller}, where we argued that {\em pseudogap} effects \cite{randeria} could lead to a bimodal homogeneous spectrum, are largely irrelevant to the spectrum of a trapped gas.
%% \cite{levtormaprivate}.  
%Trap averaging a bimodal spectrum can (in principle) lead to a unimodal spectrum.
%Furthermore, as shown by Massignan et al. \cite{massignan}, including thermal fluctuations in the $T$-matrix eliminates the bimodality which we predicted in the homogeneous spectrum in the absence of final state interactions.
%%On the other hand, these features play  a crucial role in   spatially resolved RF spectroscopy \cite{}.

Given that $n_\downarrow$ is vanishingly small, and that for short-range interactions atoms in the $\uparrow$ state will not interact with one-another, we take $n_\uparrow(r)$ to be given by the standard non-interacting result.  
\begin{equation}
n_\uparrow(r)=-\frac{1}{\lambda_T^3} g_{3/2}(-e^{\beta (\mu_\uparrow-V(r))})
\end{equation}
where we have used the Thomas-Fermi approximation to account for the slowly varying trap potential $V(r)$.   For harmonic traps, this Thomas-Fermi approximation is excellent, even at zero temperature and for small numbers of particles  \cite{muellerprofiles}.  The density is expressed in terms of the polylogarithm function $g_{3/2}(z)=\sum_j z^j/j^{3/2}$ and the thermal wavelength $\lambda_T^2=2\pi\hbar^2/mk_B T$. Similarly the down-spin density is
\begin{equation}
n_\downarrow(r)=\frac{1}{\lambda_T^3}e^{\beta [\mu_\downarrow-V(r) -\epsilon_\downarrow(n_\uparrow(r)]}
\end{equation}
where we have assumed $\mu_\downarrow<0$ and $|\beta\mu_\downarrow|\gg1$.  Defining the function $\mu(\nu)$ to be the inverse of $\nu_p$, {\em i.e.} $\nu_p[\mu(\nu)]=\nu$, and assuming a harmonic trap: $V(r)\propto r^2$, the spectrum is
\begin{equation}
I(\nu)\propto\sqrt{\mu_\uparrow^{(0)}-\mu(\nu)}e^{-\beta \epsilon_\downarrow[\mu(\nu)]}
\frac{\partial }{\partial \nu} e^{\beta \mu(\nu)}.
\end{equation}

\begin{figure}
{\psfrag{ef}{$\beta \epsilon$}
\psfrag{sm}{$\beta\mu$}
\psfrag{bm}[b]{$\beta\mu$}
\psfrag{en}[l]{$(\epsilon /n_\uparrow)\gamma$}
\includegraphics[width=\columnwidth]{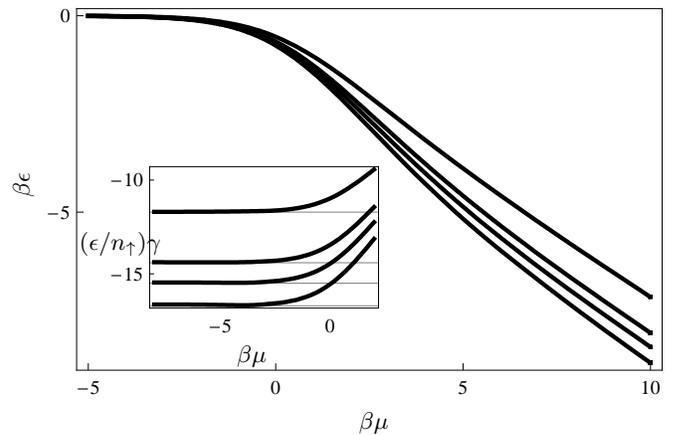}}
\caption{
Energy of a single minority species atom in a Fermi sea of majority species calculated from Nozieres and Schmidt Rink free energy.  Inset shows energy scaled by majority species density $n_\uparrow$ and the factor $\gamma=(k_b T)^{1/2}(m/\hbar^2)^{3/2}$ -- horizontal lines correspond to leading order virial expansion.  From bottom to top $a\sqrt{m k_B T/\hbar^2}=\infty,-10,-5,-2$.
}\label{epsplot}
\end{figure}

\begin{figure}[tbh]
{\psfrag{n}{$N_2$ [arb.]}
\psfrag{w}{$ \nu$ [kHz]}
\includegraphics[width=\columnwidth]{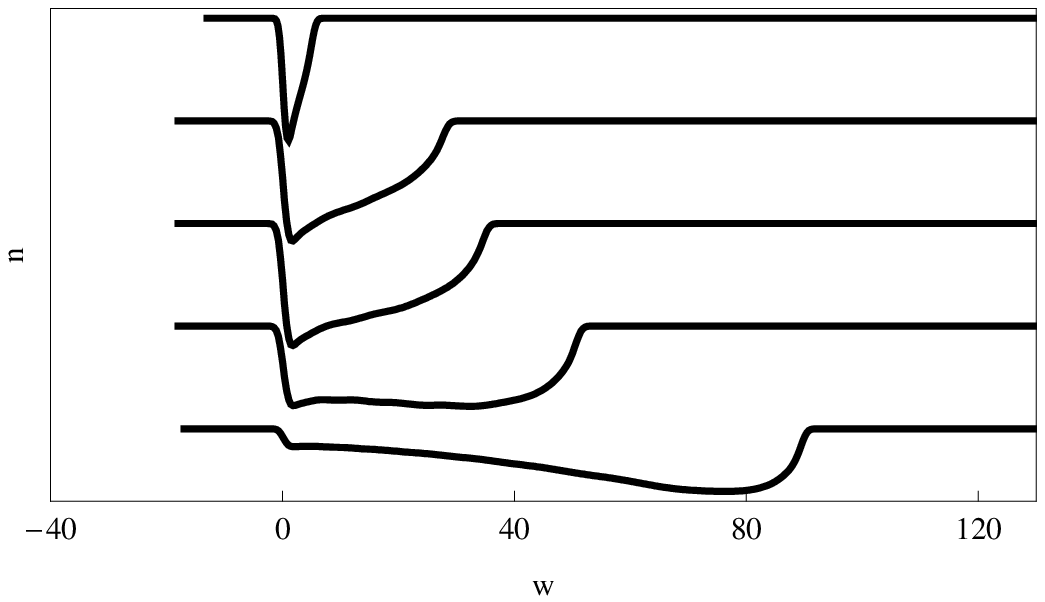}}\\[0.1in]
{\psfrag{I}{$N_2$ [arb.]}
\psfrag{w}{$h \nu/E_F$}
\includegraphics[width=\columnwidth]{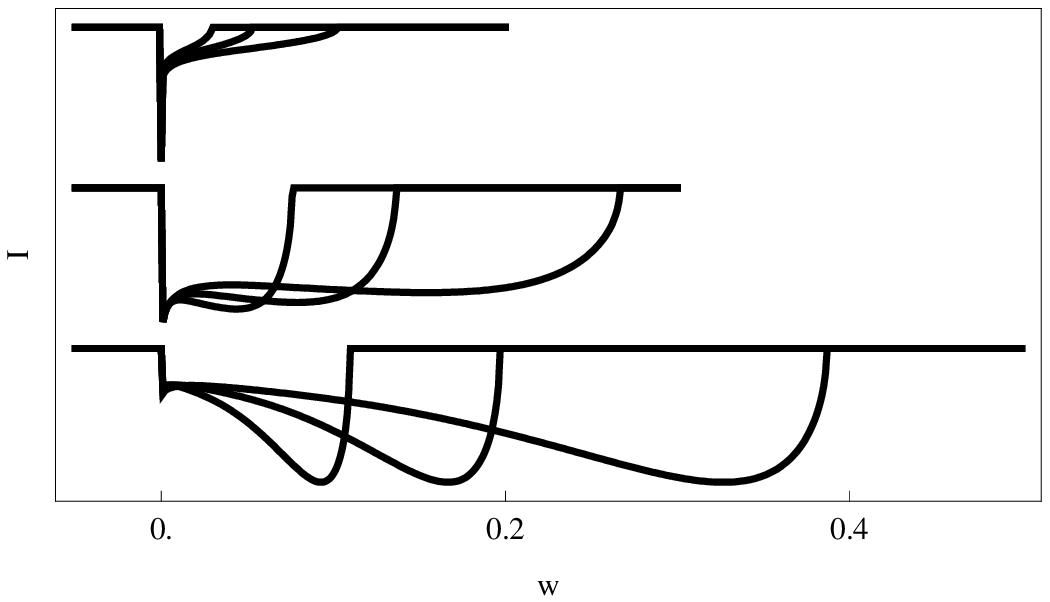}}
\caption{
Radio frequency spectra of a harmonically trapped gas in the limit of vanishing $N_2$: number of atoms remaining in the $2$ state as a function of the detuning $\nu$ of the probe from the vacuum transition value (offset vertically for clarity).  In all cases $a_{12}=\infty$.  Top figure measures frequencies in physical units for experimentally relevant parameters: from top to bottom $a_{13} \sqrt{m k_BT/\hbar^2}=-3,-2.6,-2.4,-2.4,-1.8$ with $T/T_F=1.9,1,0.9,0.7,0.5$ and $E_f=k_B T_F=260 h \mbox{kHz}, 360 h \mbox{kHz}, 360 h \mbox{kHz}, 360 h \mbox{kHz}$, where 
$E_f=(3 N \omega_x\omega_y\omega_z)^{1/3}/\pi$ measures the number of majority species atoms, with $\omega_j$ the small oscillation frequency in direction $j$, and $h$ is Planck's constant. 
Lower figure measures frequencies in units of $E_f$:  From top to bottom
$T/T_f=1.25,0.7,0.5$, while from left to right 
$a \sqrt{m k_BT/\hbar^2}=-5,-2.55,-1$.  
The shape of the spectrum is mainly set by $T/T_f$, while the frequency scale depends on the interaction strength.  The upper figures are convolved with a Gaussian of width $0.003 E_f$, while no broadening is used in the lower figures.
}\label{spec}
\end{figure}

We now argue that for short range interactions, $\epsilon_\sigma(\mu_\uparrow)$ must have the generic form illustrated in figure~\ref{epsplot}.  By considering a virial expansion, we know that for $\beta\mu_\uparrow$ sufficiently negative, $\epsilon$ must be proportional to $n_\uparrow$ \cite{hightemp}, and hence will be exponentially small in $\beta\mu_\uparrow$.  For sufficiently large $\beta\mu_\uparrow$, the scattering will be unitarity limited
and $\epsilon$ will be linear in $\beta\mu_\uparrow$.  Different theories can only change the slope of the latter relationship, and change the exact location of the crossover between these two limiting behaviors.  Regardless of these two details, the resulting inhomogeneously broadened spectrum will have the following features:
(i) At all temperatures, the edge of the cloud will contribute spectral weight at $\nu=0$:  for a harmonic trap a weak $\sqrt{-\log(\beta\nu)}$ divergence will be found.
(ii) At high temperatures ($\beta\mu_\uparrow<0,|\beta\mu_\uparrow|\gg1$),  all spectral weight will be in this peak.  Including additional sources of broadening will obscure the singularity, yielding simply an asymmetric peak.
(iii) At very low temperatures ($\beta\mu\gg1$) only an exponentially small fraction of the spectral weight will be in this low detuning peak, and it will not be observable.  Rather, most of the spectral weight will come from the center of the cloud where most of the down-spin particles reside.  
This will give a single broad peak.
At sufficiently high central density ($n_\uparrow a^3\gg1$), the location of this peak will scale
linearly with the central up-spin chemical potential.  The constant of proportionality will depend on final-state interactions.
(iv) At intermediate temperatures a bimodal structure will be seen.  Illustrative spectra are shown in figure~\ref{spec}.   It should be clear that our interpretation of the spectra is vastly different from the canonical one \cite{ketterlepairs,chingap}, where the two peaks are attributed to ``atoms" and ``pairs".

Although the qualitative spectral features are generic, we make quantitative predictions by
calculating the energies in figure~\ref{epsplot} from the generalization of  the Nozieres and Schmitt-Rink free energy \cite{citensr} to unequal densities.   Within this approximation, which can be thought of as pairwise summing all scattering events, taking into account medium effects by a phase space reduction factor,
the interaction contribution to the free energy is
\begin{equation}\label{nsr}
\begin{array}{l}
\delta\Omega=-\pi\int \frac{d^3k}{(2\pi)^3}\int \frac{d\omega}{2\pi} g(\omega) \arg\left(\frac{1}{4\pi a}+\theta(k,\omega)\right)\\
\theta(k,\omega)=\int \frac{d^3q}{(2\pi)^3}\left[\frac{1-f^\uparrow_{q-k/2}-f^\downarrow_{q+k/2}}{\omega-q^2/m-k^2/4m}+\frac{m}{q^2}\right],
\end{array}
\end{equation}
where the inverse temperature $\beta$ enters in the Bose and Fermi functions $g(\omega)=(e^{\beta\omega}-1)^{-1}$, $f^\sigma_k=(e^{\beta (k^2/2m-\mu_\sigma)}-1)^{-1}$.  Replacing $\downarrow$ with $x$ (and using the appropriate scattering length) gives the equivalent quantity in the excited state.
As in the standard derivation of the Gibbs-Duhem relation, dimensional analysis requires that the free energy density is of the form $\beta\Omega/V= \lambda^{-3}f(a/\lambda,\beta\mu_\uparrow,\beta\mu_\downarrow)$.
Since in the limit of low down-spin density this will be proportional to the number of down-spins, we have that the interaction energy will be of the form
$N_\downarrow \epsilon_\downarrow=E_{\rm int}$ $=\Omega-\Omega_0-$ $T S-\mu_\uparrow N_\uparrow - \mu_\downarrow$ $=(-3/2)\delta\Omega-(1/2) a \partial\delta\Omega/\partial a.$
  As one tunes from the dilute limit, $n a^3\ll1$, to unitarity, the interaction energy goes from $E_{\rm int}=-2\delta\Omega$ to $E_{\rm int}=-3/2\delta\Omega$ \cite{hothermo}.
Formally one can write Eq.~(\ref{nsr}) in the limit $\beta\mu\to-\infty$,
\begin{equation}\label{hitexp}
\begin{array}{l}
\left.
\frac{\delta\Omega}{N_\downarrow}\right|_{N_\downarrow=0}=
2\lambda_T^{3} \int_0^\infty
 dq\int_0^\infty d\nu\,q^2 e^{-\beta(\nu+\frac{q^2}{4m}-\mu_\uparrow)}\times\\\qquad\qquad\qquad
 \arg\left(\frac{1}{4\pi a}+\theta\right).
 \end{array}
\end{equation}
Where $\theta$, defined in Eq.~(\ref{nsr}), is evaluated with $f^\downarrow=0$, implying it is only a function of $\nu=\omega+\mu_\uparrow+\mu_\downarrow-q^2/4m,\beta,q$ and $\mu_\uparrow$.
At low temperature, $\beta \mu_\uparrow\gg1$, this integral is poorly behaved and is best replaced by a low temperature expansion whose leading term is
\begin{equation}\label{zeroT}
\left.
\frac{\delta\Omega}{N_\downarrow}\right|_{T,N_\downarrow=0}
%{\begin{array}{c}\tiny T=0\\\tiny N_\downarrow=0\end{array}}
=
\frac{1}{(2\pi)^2}\frac{\hbar^2}{m}\int_0^\infty \frac{q^2 dq}{(4\pi a)^{-1}+{\rm Re}(\theta)},
\end{equation}
where $\theta$ can be calculated analytically, and is evaluated at $\nu=q^2/4m$.
At unitarity we find $\epsilon=-0.69 \mu_\uparrow$, a reasonable approximation to the results of quantum Monte-Carlo \cite{qmc}.  Given that the lowest temperature highly polarized data from \cite{ketterlepairs} has $\beta\mu_\uparrow=1.25$, finite temperature effects are crucial and the asymptotic result cannot be compared to experiment.  

As previously anticipated, at high temperatures Eq.~(\ref{hitexp}) reduces to the polarization-imbalanced virial expansion result $\epsilon=\sqrt{2} k_B T \lambda_T^3 n_\uparrow \left(-3 b_2/4-y \partial b_2/\partial y\right)$ (cf. \cite{hothermo}),
where the second virial coefficient is $b_2=(1/\pi)\int_0^\infty dx/(1+x^2) e^{-x^2/y}$,
with $y=m k_B T a^2/\hbar^2$.   At low densities ($n a^3\ll1$) we can analytically integrate (\ref{nsr})  to recover $\epsilon=(4\pi\hbar^2/m) a n_\uparrow$, regardless of temperature.

The spectra in figure~\ref{spec} capture many features of the experiments \cite{ketterlepairs} and of more numerically demanding theories \cite{massignan,Levin-2007}.  The most successful element is that since we include final state interactions we are able to reproduce the energy scales observed in experiment.  Additionally, as already emphasized, our intermediate temperature spectra are bimodal -- however the bimodality is much much less pronounced in our theory than in the experiment.  By comparing with the calculation of Massignan, Bruun, and Stoof \cite{massignan}, we can understand this discrepency as an artifact of neglecting the width of the homogeneous spectrum -- due in part to the fact that at higher densities the initial $|\!\downarrow\rangle$ state has overlap with many $|x\rangle$ states. The intermediate detuning spectral weight (for example around 10 kHz in figure~\ref{spec}) will be reduced by this broadening, more clearly separating the spectrum into two peaks.  In experiments, finite probe duration further broadens the spectrum, and  the finite probe size and trap anharmonicities distort the low detuning peak.

Having understood the case $n_\downarrow/n_\uparrow\ll1$, we now address the balanced situation $n_\downarrow=n_\uparrow$.  It should be clear that all of our arguments about the scaling of the energy with density carry over to the unpolarized gas.  Thus, to the extent that the homogeneous spectrum consists of a single peak, one expects that the trapped spectra should have the same qualitative features that we saw in figure~\ref{spec}.  Indeed, there is very little qualitative difference between the experimental spectra of polarized and unpolarized atoms in a trap \cite{ketterlepairs}.

On the other hand, our quantitative calculations do not generalize easily to the unpolarized case.  The physics of the unpolarized gas is more complicated than that of the highly polarized gas: with increasing $n_\downarrow$ the down-spin atoms occupy a larger range of momentum states,  the up-spin Fermi sea becomes increasingly perturbed, and the normal state eventually becomes unstable to forming a superfluid.  While RF spectroscopy is undoubtedly sensitive to these many-body effects, their manifestation may be subtle, sensitive to final state interactions, and easily obscured by the trap inhomogeneities.  There does not appear to be a model independent way to use inhomogeneously broadened spectra to demonstrate superfluidity or extract a pairing gap.
% Tomographic methods \cite{Ketterle-II-2007}.  

Finally, we mention that with minor changes our qualitative arguments also apply in the simpler case of a weakly interacting Bose gas, such as spin polarized hydrogen \cite{hazzard}.  In these gases the interaction energy is proportional to the density.  At low temperature, $n\lambda_T^3>1$, the density is proportional to the chemical potential, while at high temperature, $n\lambda_T^3\ll1$, the density is exponentially small in $\beta\mu$.  Following the arguments presented here, one will therefore see a bimodal spectrum in a trapped Bose gas at intermediate temperatures.  Such bimodality is indeed found \cite{killian}.

I would like to thank 
 Sourish Basu, Francesco Fumarola,  Kaden Hazzard,
Tin-Lun Ho, Wolfgang Ketterle, Kathy Levin, Mohit Randeria,  Henk Stoof, Paivi Torma, and W. Zwerger for for various discussions and critical comments about this work.
% and ref. \cite{fumarolamueller}.
% which were instrumental in developing this work.
This work was supported in part by the National Science Foundation through grants PHY-0456261 and PHY-0758104.


\begin{thebibliography}{55}

%Molecules
\bibitem{Jin-2005} M. Greiner, C.A. Regal, and D.S. Jin, 
%Probing the Excitation Spectrum of a Fermi Gas in the BCS-BEC Crossover Regime,
Phys. Rev. Lett. 
{\bf 94}, 070403 (2005).   
%\url{http://link.aps.org/abstract/PRL/v94/e070403}


\bibitem{chingap} C. Chin, 
M. Bartenstein, A. Altmeyer, S. Riedl, S. Jochim, J. H. Denschlag, and R. Grimm, 
%Observation of the Pairing Gap in a Strongly Interacting Fermi Gas,
Science 305, 1128 (2004).
%\url{http://www.sciencemag.org/cgi/content/abstract/305/5687/1128}

%Spin Imbalanced RF
\bibitem{ketterlepairs} C. H. Schunck, Y. Shin, A. Schirotzek, M. W. Zwierlein,and W. Ketterle, 
%Pairing without superfluidity
Science {\bf 316}, 867 (2007).
% \url{http://cua.mit.edu/ketterle_group/Projects_2007/Pubs_07/schu07_pairing_0702066.pdf}

%\bibitem{lee} Alexander Seidel, Henry Fu, Dung-Hai Lee, Jon Magne Leinaas, and Joel Moore, Phys. Rev. Lett. {\bf 95}, 266405, (2005); Dung-Hai Lee, and Jon Magne leinaas, Phys. Rev. Lett. {\bf 92}, 096401 (2004).

%\bibitem{rvb}P. Fazekas and P. W. Anderson, Philos. Mag. {\bf 30}, 423 (1974).

%\bibitem{muellerfrag}
%Erich J. Mueller, Tin-Lun Ho, Masahito Ueda, and Gordon Baym, Phys. Rev. A 74, 033612 (2006).

%\bibitem{mcmillan}
%W. L. McMillan, Phys. Rev. {\bf 138}, A442 (1965).






% PREVIOUS THEORIES ON RF SPECTROSCOPY
                                                                                
\bibitem{Torma-2004} J. Kinnunen, M. Rodriguez, and P. T{\"o}rma, 
%Pairing Gap and In-Gap Excitations in Trapped Fermionic Superfluids
Science {\bf 305}, 1131 (2004).
%\url{http://www.sciencemag.org/cgi/content/abstract/305/5687/1131}


\bibitem{Ohashi-Griffin-2005} Y. Ohashi and A. Griffin, 
%Single-particle excitations in a trapped gas of Fermi atoms in the BCS-BEC crossover region,
Phys. Rev. A {\bf 72}, 013601 (2005). 
%\url{http://link.aps.org/abstract/PRA/v72/e013601}

\bibitem{Levin-2005} Yan He, Qijin Chen, and K. Levin,  
%Radio-frequency spectroscopy and the pairing gap in trapped Fermi gases
Phys. Rev. A {\bf 72}, 011602(R) (2005). 
%\url{http://link.aps.org/abstract/PRA/v72/e011602}


\bibitem{Levin-2007}Yah He, Chih-Chun Chien, Qijin Chen, and K. Levin,
%Radio Frequency Spectroscopy of Trapped Fermi Gases with Population Imbalance
Phys. Rev. A {\bf 77}, 011602(R) (2008)
%arXiv:0707.2625 (2007).

\bibitem{massignan}
P. Massignan, G. M. Bruun,  and H. T. C. Stoof, 
%Twin peaks in rf spectra of Fermi gases at unitarity in the normal phase, 
Phys. Rev. A {\bf 77}, 031601(R) (2008).
% arXiv:0709.3158 (2007).
%\url{http://arxiv.org/abs/0709.3158}


\bibitem{ketterlespec}
Gretchen K. Campbell, Jongchul Mun, Micah Boyd, Patrick Medley, Aaron E. Leanhardt, Luis G. Marcassa, David E. Pritchard, Wolfgang Ketterle, 
Science {\bf 313}, 649 (2006).

\bibitem{hazzardmueller}
Kaden R.A. Hazzard, and Erich J. Mueller, 
%Hyperfine spectra of trapped Bosons in optical lattices,  
Phys. Rev. A {\bf 76}, 063612 (2007).
%arXiv:0708.3657 (2007).
%url{http://arxiv.org/abs/0708.3657}




\bibitem{Zwerger-2007} M. Punk and W. Zwerger,
%Theory of RF-spectroscopy of strongly interacting Fermions. 
arXiv:0707.0792 (2007).
%\url{http://arxiv.org/abs/0707.0792}


\bibitem{Baym-2006} Z. Yu and G. Baym, 
%Spin-correlation functions in ultracold paired atomic-fermion systems: Sum rules, self-consistent approximations, and mean fields,
Phys. Rev. A {\bf 73}, 063601 (2006). 
%\url{http://link.aps.org/abstract/PRA/v73/e063601}

\bibitem{Baym-2007} G. Baym 
, C.~J. Pethick, Z. Yu, and M.W. Zwierlein, 
%Coherence and clock shifts in ultracold Fermi gases with resonant interactions,
arXiv:0707.0859.
%\url{http://arxiv.org/abs/0707.0859}

\bibitem{Chin-Julienne-2005} C. Chin and P.S. Julienne, 
%Radio-frequency transitions on weakly bound ultracold molecules,
Phys. Rev. A {\bf 71}, 012713 (2005). 
%\url{http://link.aps.org/abstract/PRA/v71/e012713}




\bibitem{perali-RF}A. Perali, P. Pieri, G.C. Strinati, 
%Competition between final-state and pairing-gap effects in the radio-frequency spectra of ultracold Fermi atoms, 
arXiv:0709.0817 (2007).
%\url{http://arxiv.org/abs/0709.0817}

\bibitem{Levin-2008}
Yan He, Chih-Chun Chien, Qijin Chen, and K. Levin
%Temperature and final state effects in radio frequency spectroscopy experiments on atomic Fermi gases
 arXiv:0804.1429 (2008).

\bibitem{newstoof}
P. Massignan, G. M. Bruun, and H. T. C. Stoof
% Spin polarons and molecules in strongly-interacting atomic Fermi gases
 arXiv:0805.3667 (2008)
 
 
 \bibitem{sheehy}
Martin Veillette, Eun Gook Moon, Austen Lamacraft, Leo Radzihovsky, Subir Sachdev, D.E. Sheehy
arXiv:0803.2517 (2008).

%
%\bibitem{superfluid} K. M. O'Hara \textit{et al.}, Science
%\textbf{298}, 2179 (2002); C. A. Regal, Nature \textbf{424}, 47
%(2003); M. Greiner, C. A. Regal, and D. S. Jin, Nature \textbf{426},
%537 (2003); S. Jochim \textit{et al.}, Science \textbf{302}, 2101
%(2003); M. W. Zwierlein \textit{et al.}, Phys. Rev. Lett.
%\textbf{91}, 250401 (2003); C. A. Regal, M. Greiner and D. S. Jin,
%Phys. Rev. Lett. \textbf{92}, 040403 (2004); M. W. Zwierlein
%\textit{et al.}, Phys. Rev. Lett. \textbf{92}, 120403 (2004); J.
%Kinast \textit{et al.}, Phys. Rev. Lett. \textbf{92}, 150402 (2004);
%M. Bartenstein \textit{et al.}, Phys. Rev. Lett. \textbf{92}, 203201
%(2004); T. Bourdel \textit{et al.}, Phys. Rev. Lett. \textbf{93},
%050401 (2004); C. Chin \textit{et al.}, Science \textbf{305}, 1128
%(2004); M. W. Zwierlein \textit{et al.}, Nature \textbf{435},
%1047-1051 (2005); G. B. Partridge \textit{et al.}, Phys. Rev. Lett.
%\textbf{95}, 020404 (2005); J. Kinast \textit{et al.}, Science
%\textbf{307}, 1296 (2005).

%\bibitem{FF1}
%U. Fano,  Nuovo Cimento, \textbf{12},  156 (1935); Phys. Rev. A
%\textbf{ 124},  1866 (1961).

%\bibitem{FF2}
% H. Feshbach,
% Ann. Phys. \textbf{5},
% 357 (1958).

%\bibitem{review} For review and a comprehensive list of references see D. S. Petrov, C. Salomon, and
%G. V. Shlyapnikov, J. Phys. B 38, S645 (2005); V. Gurarie and L.
%Radzihovsky, cond- mat/0611022.
%
%\bibitem{hulet} G. B. Partridge \textit{et al.}, Science, \textbf{311}, 503 (2006).

%\bibitem{ketterle} M. W. Zwierlein \textit{et al.}, Science, \textbf{311}, 492 (2006);
%Nature, \textbf{442}, 54 (2006).

%\bibitem{Cl} A. M. Clogston, Phys. Rev. Lett. \textbf{9}, 266 (1962).

%\bibitem{Ch} B. S. Chandrasekhar, Appl. Phys. Lett. \textbf{1}, 7 (1962).

%\bibitem{pairs}
%C. A. Regal, C. Ticknor, J. L. Bohn, D. S. Jin, Nature 424, 47 (2003).

%

%% C. Regal, D. Jin, Phys. Rev. Lett. 90, 230404(2003);

%%C. Chin and P. Julienne
%%Phys. Rev. A 71,012713 (2005),  cond-mat/0408254

%

%\bibitem{chin} C.Chin \emph{et al.}, Science \textbf{305}, 1128 (2004).

%\bibitem{Schunck}  C. H. Schunck \emph{et al.}, Science \textbf{316}, 867 - 870.

%\bibitem{gupta} S. Gupta \emph{et al.}, Science \textbf{300}, 1723 (2003).

%\bibitem{Zoller} G. M. Bruun \emph{et al.}, Phys. Rev. A
%\textbf{64}, 033609.

%\bibitem{Griffin} Y. Ohashi and A. Griffin, Phys. Rev. A \emph{72}, 013601
%(2005).

%\bibitem{kinnunen} J. Kinnunen, M. Rodriguez, P. Torma, Phys. Rev. Lett. \textbf{92}, 230403 (2004).

%\bibitem{kohn} W. Kohn, Phys. Rev. Lett. \textbf{2}, 393 (1959).

%
%\bibitem{yacoby} O. Auslaender, A. Yacoby, R. de Picciotto, K. W. Baldwin, L. N. Pfeiffer, and K. W. West, %ñTunneling Spectroscopy of the Elementary Excitations in a One-Dimensional Wireî,
%Science 295, 825-828 (2002).

%\bibitem{Baym} Y. Zu and G. Baym, Phys. Rev. A \textbf{73}, 063601
%(2006).

%\bibitem{levin} Q. Chen,
%et. al., Phys. Rep. 412, 1-88 (2005).

%\bibitem{Strinati} Phys. Rev. B \textbf{72}, 012506 (2005).

%
%\bibitem{Combescot1} R. Combescot, X. Leyronas and M. Yu Kagan,
%Phys. Rev. A \textbf{73}, 023618 (2006).

%\bibitem{Combescot2} R. Combescot \emph{et al.},
%cond-mat/0702314v1.

%%\bibitem{fumarola} F. Fumarola, I.L. Aleiner, and B.L. Altshuler, arXiv:cond-mat/0703003v2.

%
%\bibitem{kadanoffmartin}  L.P. Kadanoff and P.C. Martin, Phys. Rev. \textbf{124}, 670 (1961).

%%\bibitem{combescot3} R. Combescot, X. Leyronas and M. Yu. Kagan,
%%Phys. Rev. A \textbf{73}, 023618 (2006).


% EXPERIMENTS ON RF SPECTROSCOPY

%Original RF spec
%\bibitem{Grimm-2004} C. Chin, M. Bartenstein, A. Altmeyer, S. Riedl, S. Jochim, J. Hecker Denschlag, and R. Grimm, Science {\bf 305}, 1128 (2004).
% see chingap
 
% \bibitem{fumarolamueller}
% Francesco Fumarola, Erich J. Mueller
%%  Single particle spectrum of resonant population imbalanced Fermi gases
 % arXiv:0706.1205 (2006).
%%\url{http://arxiv.org/abs/0706.1205}



\bibitem{basumueller}
Sourish Basu and Erich J. Mueller, Phys. Ref. Lett. {\bf 101}, 060405 (2008).


\bibitem{insensitive}
S. Gupta et al.
%, Z. Hadzibabic, M. W. Zwierlein, C. A. Stan, K. Dieckmann, C. H. Schunck, E. G. M. van Kempen, B. J. Verhaar, and W. Ketterle, 
Science {\bf 300}, 475 (2003);
M.W. Zwierlein et al. 
%Z. Hadzibabic, S. Gupta, and W. Ketterle,
Phys. Rev. Lett. {\bf 91}, 250404 (2003).

%Spatial Resolved RF
\bibitem{Ketterle-II-2007} Y. Shin 
, C.H. Schunck, A. Schirotzek, and W. Ketterle, 
%Tomographic rf Spectroscopy of a Trapped Fermi Gas at Unitarity,
Phys. Rev. Lett. {\bf 99}, 090403 (2007).
%\url{http://link.aps.org/abstract/PRL/v99/e090403}

\bibitem{Ketterle-2008}
Christian H. Schunck, Yong-il Shin, Andre Schirotzek, Wolfgang Ketterle,
%Determination of the Fermion Pair Size in a Resonantly Interacting Superfluid
Nature 454, 739-743 (2008).
% arXiv:0802.0341 

\bibitem{pethstoof} C. J. Pethick and H. T. C. Stoof, Phys. Rev. A 64, 013618 (2001)

\bibitem{mahan} %For a particularly simple case of a static impurity, the calculation appears in 
Gerald D. Mahan, ``Many Particle Physics", Plenum Press (New York, 1990), chap.4.1.





%\bibitem{ketnew}
% Y. Shin, M. W. Zwierlein, C. H. Schunck, A. Schirotzek, and W. Ketterle
%Phys. Rev. Lett. 97, 030401 (2006)


%\bibitem{campbell}
%G.K. Campbell et al.
%, J. Mun, M. Boyd, P. Medley, A.E. Leanhardt, L. Marcassa, D.E. Pritchard, W. Ketterle,
%Imaging the Mott Insulator Shells By Using Atomic Clock Shifts 
%Science 313 , 649-652 (2006).
%\url{http://www.sciencemag.org/cgi/content/abstract/313/5787/649}










%\bibitem{randeria}
%C. A. R. S‡ de Melo, Mohit Randeria, and Jan R. Engelbrecht, Phys. Rev. Lett. 71, 3202 - 3205 (1993);
%Mohit Randeria in Proceedings of the International School of Physics ``Enrico Fermi''
%Course CXXXVI on High Temperature Superconductors'
%edited by G. Iadonisi, J. R. Schrieffer, and M. L. Chiafalo, (IOS Press,
%1998), p.53 - 75; cond-mat/9710223.


%\bibitem{levtormaprivate} This was pointed out in separate private communications with Kathy Levin and Paivi Torma.

\bibitem{muellerprofiles}
Erich J. Mueller, Phys. Rev. Lett. {\bf 93},190404 (2004).


\bibitem{hightemp}
Tin-Lun Ho, and Erich J Mueller, 
%High Temperature Expansion Applied to Fermions near Feshbach Resonance, 
Phys. Rev. Lett. {\bf 92}, 160404 (2004).
%\url{http://link.aps.org/abstract/PRL/v92/e160404}


\bibitem{citensr}P. Nozi\`eres and S. Schmitt-Rink, J. Low Temp. Phys. {\bf 59}, 195 (1985). 

\bibitem{hothermo}
Tin-Lun Ho, Phys. Rev. Lett. 92, 090402 (2004).

\bibitem{qmc}
C. Lobo, A. Recati, S. Giorgini, and S. Stringari, 
Phys. Rev. Lett. 97, 200403 (2006).

\bibitem{hazzard} Kaden R.A. Hazzard, private communications.

\bibitem{killian} Dale G. Fried,
Thomas C. Killian, Lorenz Willmann, David Landhuis, Stephen C. Moss, Daniel Kleppner, and Thomas J. Greytak,
 Phys. Rev. Lett. {\bf 81}, 3811 (1998).

\end{thebibliography}
\end{document}